\documentclass[12pt]{iopart}
\pdfoutput=1
\usepackage{color}
\usepackage{graphicx}
\usepackage{subfigure}
\usepackage{lineno}

\usepackage{latexsym,amsmath,verbatim}
\usepackage{iopams}

\begin{document}

\title{Graph theoretical approaches for the characterization of damage in hierarchical materials}
\author{Paolo Moretti, Jakob Renner, Ali Safari and Michael Zaiser}
\address{\address{Dept. of Materials Science, WW8-Materials Simulation, FAU Universit\"at Erlangen-N\"urnberg, Dr.-Mack-Stra{\ss}e 77, 90762 F\"urth, Germany}}

\begin{abstract} 
We discuss the relevance of methods of graph theory for the study of damage in simple model materials described by the random fuse model. While such methods are not commonly used when dealing with regular random lattices, which mimic disordered but statistically homogeneous materials, they become relevant in materials with microstructures that exhibit complex multi-scale patterns. We specifically address the case of hierarchical materials, whose failure, due to an uncommon fracture mode, is not well described in terms of either damage percolation or crack nucleation-and-growth. We show that in these systems, incipient failure is accompanied by an increase in eigenvector localization and a drop in topological dimension. We propose these two novel indicators as possible candidates to monitor a system in the approach to failure. As such, they provide alternatives to monitoring changes in the precursory avalanche activity, which is often invoked as a candidate for failure prediction in materials which exhibit critical-like behavior at failure, but may not work in the context of hierarchical materials which exhibit scale-free avalanche statistics even very far from the critical load.   
\end{abstract}


\maketitle

\section{Introduction}

\begin{figure}
\begin{center}
\vspace{-0cm}
\includegraphics[width=13cm]{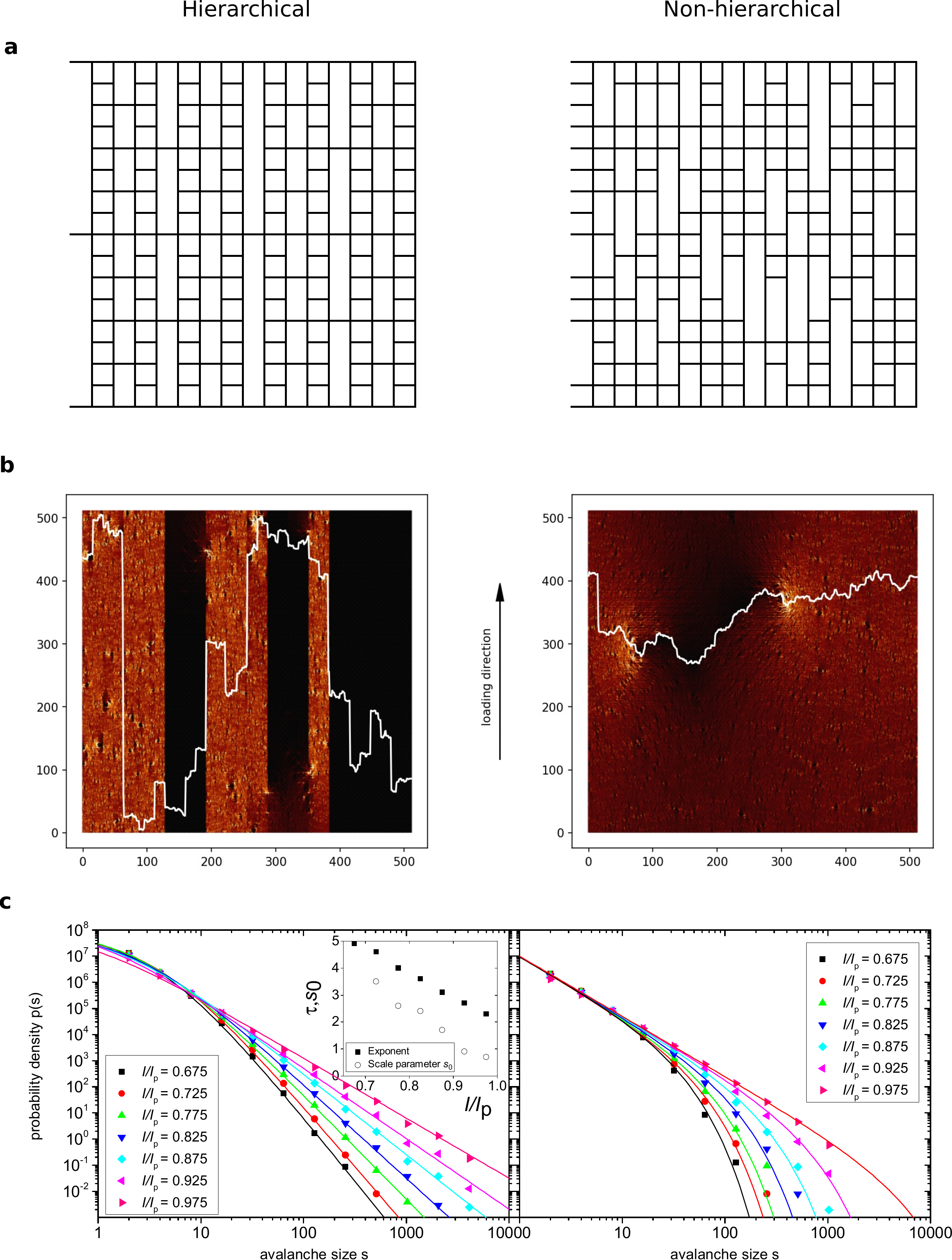}
\caption{Comparison between hierarchical fuse networks (HFN) and reference random fuse networks (RFN), with the same number of horizontal links. Row a: Structures of HFN and RFN, for systems of size $L=16$. Row b: comparisons between current profiles (colored sites) close to the peak current $I_p$, and between final crack profiles (while lines), for systems of size $L=512$. Row c: Avalanche size distributions for HFN and RFN, for values of $I$ gradually approaching $I_p$, for systems of size $L=512$.}
\label{fig:model}
\end{center}
\end{figure}

Materials with hierarchical microstructures are characterized by microstructure patterns that repeat on different length scales in a self-similar fashion. This type of organization is encountered in nature in systems such as collagen \cite{Gautieri2011_NL}, bone \cite{Rho1998,Gupta2006_PNAS}, wood \cite{Fratzl2007_PMS}, and nacre \cite{Sun2012_RSCA,Jiao2015_SCREP}, and is believed to provide enhanced toughness, contain damage and impede crack propagation. Recent numerical results for hierarchical fuse networks confirm that the hierarchical organization is responsible for the systematic confinement of fracture patterns \cite{Moretti2018_SCREP}. The resulting crack profiles differ from the self-affine scaling behavior that is found in experimental studies for a wide range of materials from metals and glasses\cite{Ponson2006_PRL,Bonamy2006_PRL} to geomaterials \cite{Ponson2007_PRE} and paper \cite{Salminen2003_EPJB} as well as in theoretical investigations \cite{Zapperi2005_PRE,Alava2006_AP} which represent fracture using random fuse network (RFN) or random beam network models. In hierarchical materials, by contrast, the fluctuations of the fracture surface profile cannot be described in the sense of stochastic roughening and quantified by standard roughness exponents. Such materials can be modelled in terms of hierarchical fuse networks (HFN) or hierarchical beam networks. For different methods of constructing such networks, see \cite{Moretti2018_SCREP}. In HFN, the fracture surfaces exhibit discontinuities in the form of abrupt jumps which are power law distributed in size. At the same time, in such hierarchical systems the fracture precursor activity which characterizes damage accumulation before failure exhibits scale-invariant behavior at every loading stage. In non-hierarchical RFN, microcrack accumulation proceeds in avalanches whose power law distribution has an exponential cut-off which diverges at the peak current $I_\textrm{p}$. This allows one to envisage failure as a critical phenomenon and large avalanches as precursors of failure. In HFN, by contrast, scale-invariant avalanche-size distributions are encountered well before $I_\textrm{p}$, suggesting that the statistical signatures of damage accumulation do not undergo any qualitative changes as one approaches the peak load. The same is true for the post-peak behavior in fracture simulations that are carried out under voltage control: also in this regime, microcracking of HFN occurs in bursts which continue to exhibit power law size distributions and no apparent cut-off. We may thus speak of generic scale-invariant behavior that, unlike critical behavior, is not contingent upon fine tuning the system to a critical point (in RFN: the peak load $I_p$). Figure \ref{fig:model} illustrates the difference between RFN - mimicking statistically homogeneous disordered materials - and HFN mimicking hierarchically structured materials, both in terms of the load dependent avalanche statistics and the characteristic damage and stress pattern during failure. 

The observation of generic scale invariant behavior in hierarchically patterned networks is by no means new. Activity propagation in the brain, which also exhibits a hierarchical structure, is in fact believed to benefit from such generic scale invariance in dynamic behavior \cite{Kaiser2007,Moretti2013}. In the case of materials, generic scale invariance may however imply that the precursor activity cannot be easily used to assess the proximity to failure. The behavior is in fact analogous to the properties of percolation transitions in hierachical networks. In this case too, hierarchical networks do not display a clear percolation threshold \cite{Boettcher2009,Friedman2013}: upon removing a fraction $p$ of links, power law distributions of connected component sizes are encountered for every value of $p$, suggesting that connected component statistics is not a suitable tool for assessing the state of damage, and even less so, the resilience of a hierarchical material. 

Here we aim at filling the conceptual gap that results from the inadequacy of standard methods for the study of damage in hierarchical materials. To do so, we resort to methods borrowed from network science and graph theory. The idea of using networks to model materials is of course not new. Networks have been used in the past to describe force chains e.g. in granular materials or contact patterns in colloidal dispersions. Models of hierarchical materials may benefit from recent advances in the application of concepts of spectral theory and topology to the study of hierarchical networks. In the following we will focus in particular on the role of eigenvalue localization and Hausdorff dimension, which we believe may provide better indicators of the state of damage in such systems than conventional indicators such as divergence of avalanche sizes. 

\section{Hierarchical fiber networks}
Hierarchical fiber network models are characterized by two main features: i) a modular organization, where modules are densely connected components which exhibit weaker mutual connections; ii) a hierarchical encapsulation, where smaller modules are embedded into larger less connected modules, in an iterative fashion. It was shown that hierarchical fuse networks (HFN) of this type exhibit generic critical-like behavior, regardless of their deterministic or randomized nature \cite{Moretti2018_SCREP}. The distinctive feature of a sheet material with hierarchical load carrying structure is that the material is, parallel to the loading direction, sub-divided by cuts of power law distributed lengths, which partition the sample into load carrying modules with power law distributed sizes. Example structures of this type are shown in Figure \ref{fig:model}, details of the construction process are discussed in Ref. \cite{Moretti2018_SCREP}.

From the point of view of graph theory each elementary segment in Figure \ref{fig:model}a is an edge of the graph. Each edge is delimited by two nodes. In the simple case in Figure \ref{fig:model} nodes are laid out as in a 2D square lattice, whereas more complex and higher dimensional arrangements are possible. We call $N$ the total number of nodes, $N_x$ ($N_y$) the number of nodes in the horizontal (vertical) direction, and $E$ the number of edges. The HFN in Figure \ref{fig:model} is periodically continued in the horizontal direction, and its linear size in both the $x$ and the $y$ direction is $L=N_x$ (in adimensional units). As for the vertical direction, the top and bottom rows act as boundaries. According to the specific construction implemented in Figure \ref{fig:model}, $N_y=N_x+1$ and the total number of non-boundary nodes is $N'=N-2N_x=N_x(N_x-1)$.   

To simulate damage in these structures, we follow the standard quasi-static simulation method for the random fuse model (RFM). We consider each edge as a load carrying element of unit conductivity. Voltages $V_i$ can be measured at each node $i$ and result in a current $I_{ij}=V_i-V_j$ carried by edge $ij$. Boundary conditions are applied as fixed voltages at the top boundary nodes, while each link $ij$ is initially assigned a fixed random current threshold $t_{ij}$. In each iteration of the simulation, the boundary voltage is set to $V=1$, and equilibrium voltages at each non-boundary node are calculated solving numerically the Kirchhoff nodes' law equations (algebraic sum of incident currents is zero) for each node. Currents are computed accordingly. The edge carrying the highest $r=I_{ij}/t_{ij}$ will be the first one to break as soon as the external $V$ is set to $1/r$. Consequently, the weak edge is removed, the global $I$ and $V$ (correctly rescaled by $1/r$) are stored, and the simulation moves to the next iteration, until formation of a system wide crack sets the global conductivity to zero. While results generally depend on the choice of the probability distribution function $P(t_{ij})$, it was shown that the qualitative features of the damage accumulation process, viz., the emergence of super-rough cracks and the observation of generic scale invariance in the precursor statistics, are not affected by changing  $P(t_{ij})$ from a uniform distribution (widely used in the statistical physics literature) to a Weibull distribution (more realistic, according to reasoning in terms of materials strength) \cite{Moretti2018_SCREP}. The numerical results shown in the following consider the case of a uniform threshold distribution, for the deterministic HFN introduced above, but the findings carry over to Weibull distributed thresholds.

\section{Spectral theory and localization}\label{sec:localization}           
In this section we introduce the spectral graph analysis for fuse networks and highlight its main results for the HFN models considered. To this end, we will introduce two different matrices: i) the Laplacian matrix $\mathrm{L}$, which appears in several graph theoretical studies; and ii) the Dirichlet Laplacian $\mathcal{L}$, which arises from $\mathrm{L}$ when applying boundary conditions and provides the correct description of our RFM equations. We recall that the Laplacian nature of the RFM equilibrium equations can be seen by writing the generic Kirchhoff nodes' law for each node $i$ as
\begin{equation}\label{eq:Kirchhoff}
\sum_{i=1}^N \mathrm{A}_{ij}(V_i-V_j)=0,
\end{equation}
where $\mathrm{A}_{ij}$ is the generic element of the symmetric $N\times N$ adjacency matrix $\mathrm{A}$: by definition $\mathrm{A}_{ij}=1$ if $i$ is neighbour of $j$, or zero otherwise. Eq. \ref{eq:Kirchhoff} can be rewritten as
\begin{equation}\label{eq:Laplace}
\mathrm{L}\mathbf{V}=0,
\end{equation}
where $\mathbf{V}$ is the column vector  of generic element $V_i$ and $\mathrm{L}$ is the Laplacian matrix of the network (more precisely, the symmetric non-normalized graph Laplacian), with $\mathrm{L}_{ij} = \delta_{ij}\sum_l \mathrm{A}_{jl} -\mathrm{A}_{ij}$ .  Eq. \ref{eq:Laplace} simply states that the equilibrium equation for our scalar elasticity is a discretized Laplace equation of the type $-\nabla^2\mathbf{V} = 0$. By construction, if the network consists of a unique connected component, the Perron-Frobenius theorem ensures that $\mathrm{L}$ has eigenvalues $0=\mu_1<\mu_2\le \mu_3 ...$. Being $\mu_1=0$, Eq. \ref{eq:Laplace} has infinitely many solutions, arising from the (unique) zero eigenvalue $\mu_1$. This property simply reflects the fact that infinitely many equilibrium states are possible when no boundary conditions are applied.  By applying boundary conditions (i.e. by fixing the values of $V_i$ for $1\le i \le N_x$ and $N-N_x+1 \le i \le N$), Eq. \ref{eq:Laplace} is reduced to a $N^\prime-$dimensional system of equations of the form 
\begin{equation}\label{eq:reduced}
\mathrm{\mathcal{L}}\mathbf{v}=\mathbf{b},
\end{equation}            
where we recall that $N'=N-2N_x$ is the number of non-boundary nodes. Here $\mathrm{\mathcal{L}}$ is a Dirichlet Laplacian, with lowest eigenvalue $\lambda_1>0$, which ensures a unique solution to the electric/elastic equilibrium problem. Here $\mathbf{v}$ is a vector whose $v_i$ element is the voltage at non-boundary node $i$, while $\mathbf{b}$ is fixed and depends on the voltages at the boundary nodes. 

\begin{figure}
\begin{center}
\vspace{-0cm}
\includegraphics[width=14cm]{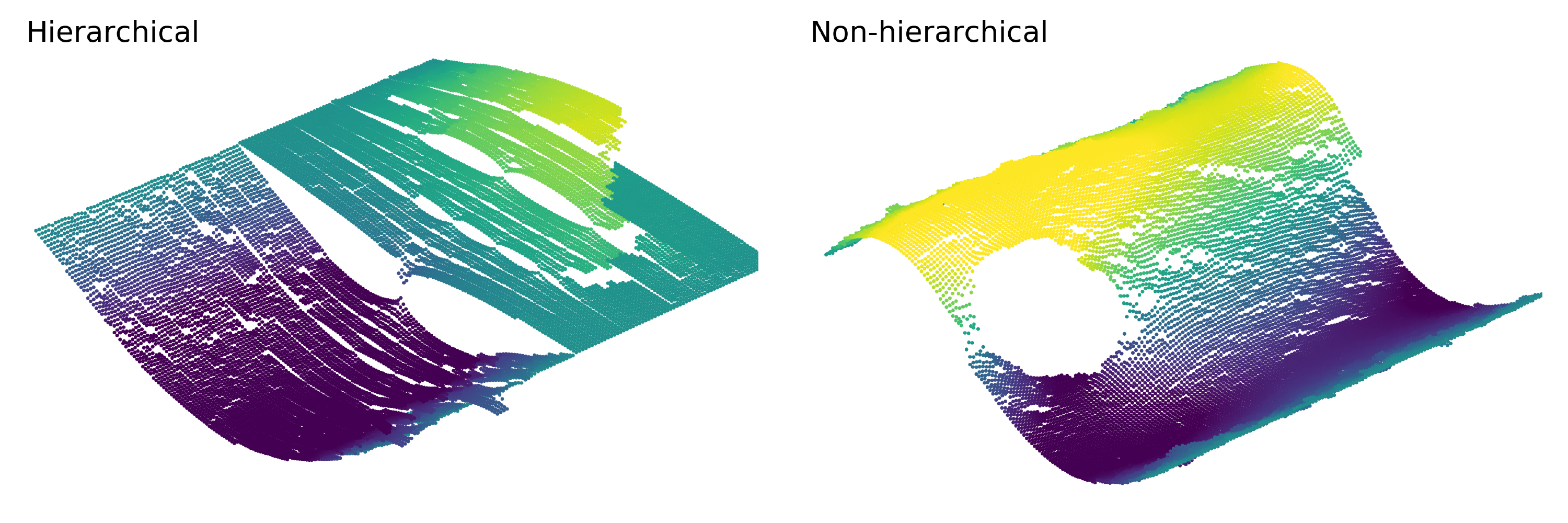}
\caption{Two eigenvectors in the lower spectral edge of the Dirichlet Laplacian for a HFN (left) and a reference square lattice (right) close to the peak current.  Such eigenvectors act as building blocks of the voltage profile in Eq. \ref{eq:solution}. In the case of HFN, the eigenvector is localized, meaning that a significant subset of it is close to zero (a green flat surface in figure). In the square lattice case instead, the eigenvector is trivially a plane wave, as expected for a periodic system. The individual eigenvectors point to two substantially different fracture modes: scattered microcracks in the hierarchical case, a single critical crack in the square lattice. In both cases, the eigenvectors corresponding to the second lowest eigenvalues of $\mathcal{L}$ are plotted, although similar results would be obtained for any choice of eigenvectors in the respective lower spectral edges.}
\label{fig:eigenvectors}
\end{center}
\end{figure}

While Eq. \ref{eq:reduced} represents the normal form in which the equilibrium problem is fed to a numerical solver in order to compute $\mathbf{v}$, here we calculate its analytical solution, which can be formally expressed in terms of the eigenvalues $\lambda_1, \lambda_2, ... $ and corresponding normalized eigenvectors $\mathbf{w}_1, \mathbf{w}_2, ...$ of $\mathrm{\mathcal{L}}$ as
\begin{equation}\label{eq:solution}
\mathbf{v}=\sum_{m=1}^{N^\prime} \frac{1}{\lambda_m} (\mathbf{w}_m \cdot \mathbf{b})\mathbf{w}_m,
\end{equation}
where $ \cdot $ indicates the standard scalar product. It can be seen from Eq. \ref{eq:solution} that the equilibrium configuration can be expressed as a linear combination of eigenvectors of $\mathrm{\mathcal{L}}$, in which eigenvectors corresponding to near-zero eigenvalues (the {\it lower spectral edge}) have a high weight because of the $1/\lambda_m$ factor. While these considerations may be of little relevance for regular lattices, which do not exhibit exotic spectral properties, they become crucial in hierarchical networks like the ones considered here. In particular, hierarchical modular networks: (i) possess a number of exceedingly small eigenvalues; (ii) these eigenvalues are associated with localized eigenvectors \cite{Moretti2013}. Eigenvector localization \cite{Goltsev2012,PastorSatorras2016_SCREP} is the property by which the components of an eigenvector are all vanishing except for a subset (see Fig. \ref{fig:eigenvectors} for an eigenvector of a HFN). Property (ii) tells us that the building blocks of the final $\mathbf{v}$ are localized patches in the networks, while property (i) ensures that only a few of those patches contribute to $\mathbf{v}$, and thus that voltages/displacements and currents/loads are localized in specific regions of the system. Localization of loads implies that, rather than growing a critical crack, the system develops localized patterns of damage, which eventually coalesce into a super-rough crack as the one shown in Fig. \ref{fig:model}.

\begin{figure}
\begin{center}
\vspace{-0cm}
\includegraphics[width=14cm]{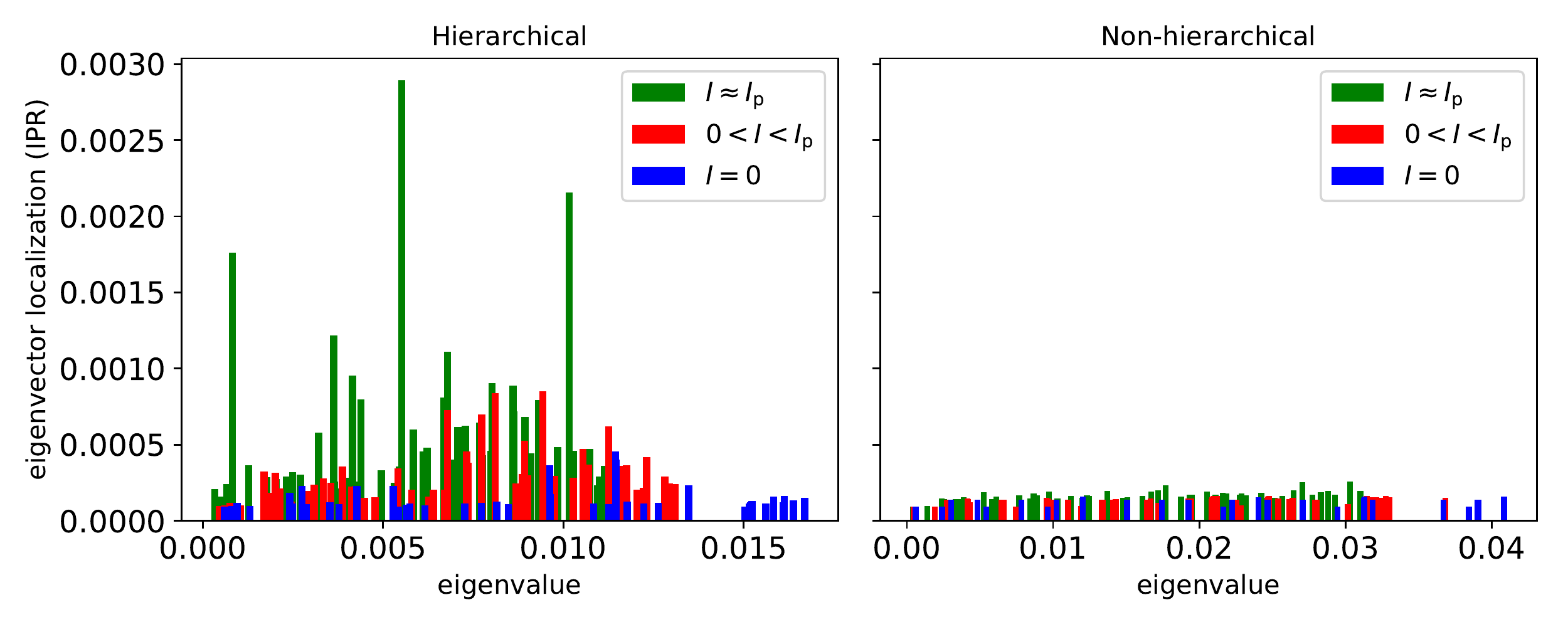}
\caption{Eigenvector localization in a HFN (left) and in a reference square lattice (right), at three different load stages. The systems are the same as the ones shown in Figure \ref{fig:eigenvectors}. The $x$ coordinate of each bar indicates an eigenvalue in the lower spectral edge of the Dirichlet Laplacian, while its height measures the inverse participation ratio of its corresponding eigenvector. Incipient failure is accompanied by an increase in localization in both cases, but significantly more strongly for the HFN.}
\label{fig:iprs}
\end{center}
\end{figure}

In order to quantify eigenvector localization, it is customary to consider the inverse participation ratio (IPR) for the $m-$th eigenvector as 
\begin{equation}\label{eq:ipr}
I_m = \frac{ \sum_i (w_{mi})^4 }{ \sum_i (w_{mi})^2 },
\end{equation}
with higher values of $I_m$ pointing to stronger localization \cite{Goltsev2012}.  Fig. \ref{fig:iprs} shows values of the IPR for eigenvectors corresponding to the $20$ lowest eigenvalues, for a HFN and for a reference square lattice, at three different stages of the loading curve. While our results confirm the intuition of an increase of localization near the peak load, such an increase is much more significant in the hierarchical case, where it nears a one-order-of-magnitude jump in IPR at $I_p$.  

\begin{figure}
\begin{center}
\vspace{-0cm}
\includegraphics[width=7cm]{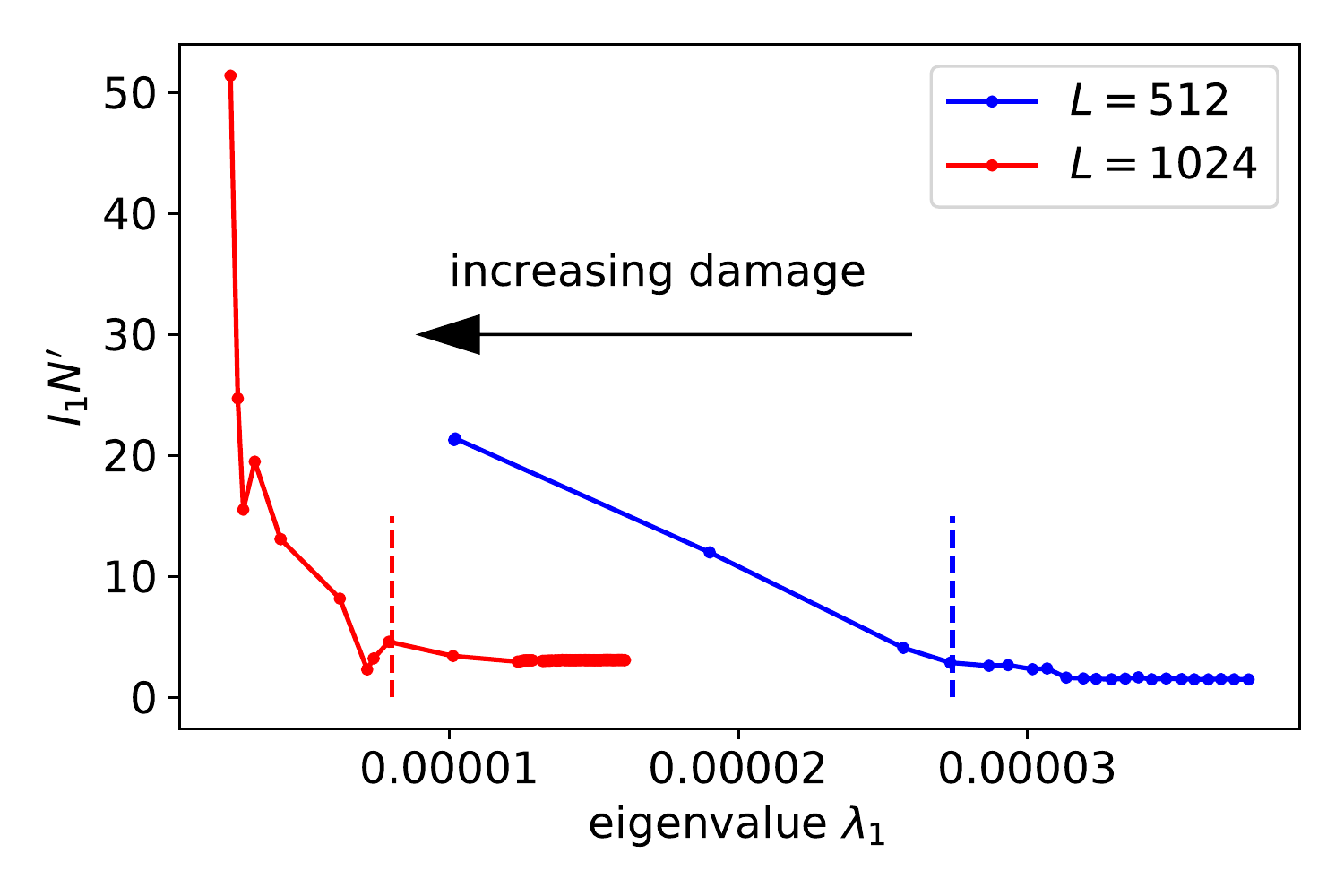}
\includegraphics[width=7cm]{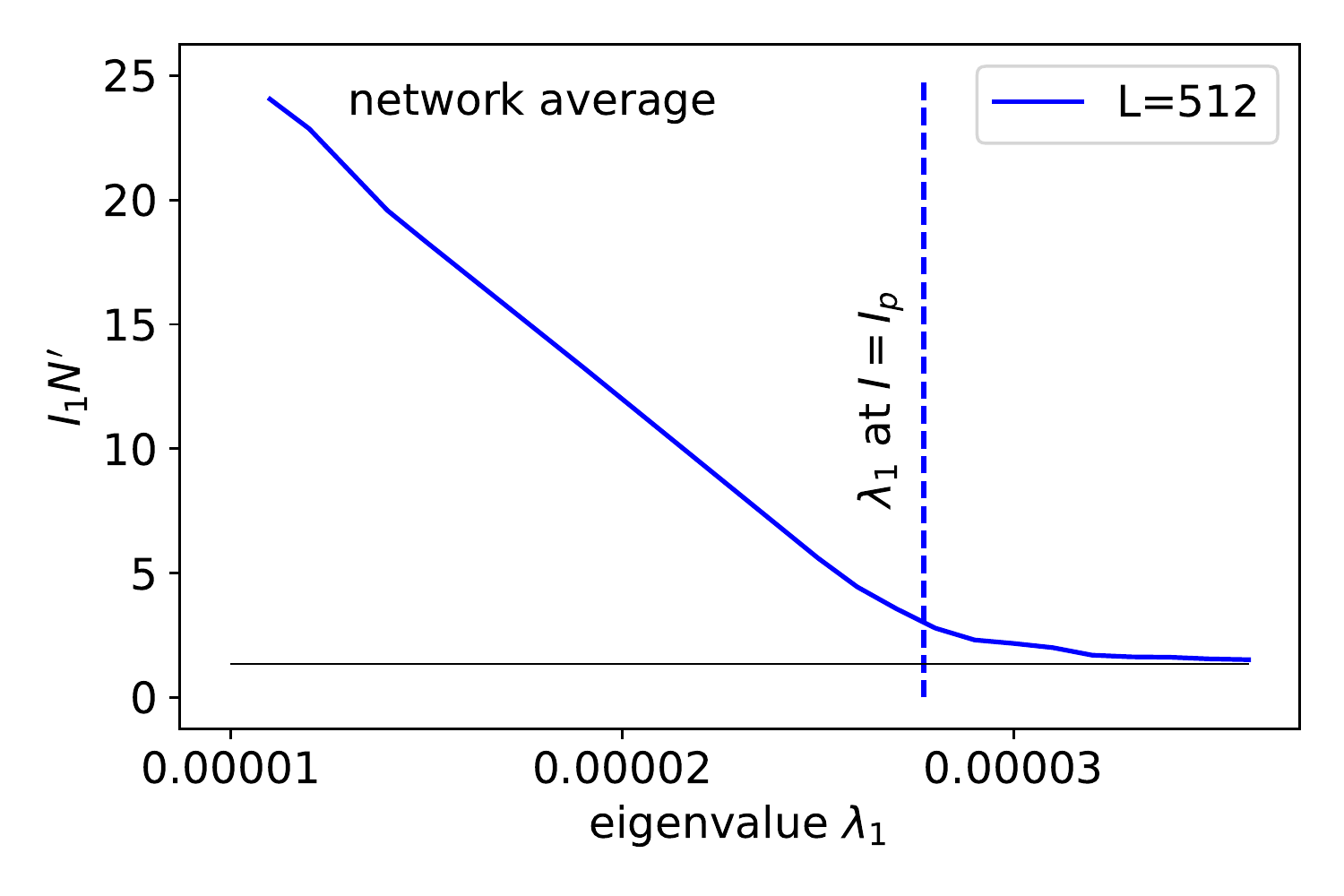}
\caption{Evolution of localization in large-scale HFN under load. Curves should be read from right (undamaged system) to left (cracked system). For each curve, the data corresponding to the system at peak stress $I_p$ is marked by a vertical dashed line of the same color. Left: individual realizations of systems of size $L=512$ and $L=1024$. Right: average over $5$ network realization, showing the substantial robustness of the results on the left.}
\label{fig:ipr_increase}
\end{center}
\end{figure}

A clearer picture of how effectively localization signals the vicinity to $I_p$ in HFN is provided in Figure \ref{fig:ipr_increase} for much larger system sizes. The inverse participation ratio determined for an eigenvector in the lower spectral edge (in this case, the eigenvector corresponding to the lowest eigenvalue) starts to increase exactly at the peak current, providing a clear-cut indicator of the system's incipient failure. Figure \ref{fig:ipr_increase} (left) also provides important information regarding the size dependence of damage induced localization, as we detail in the following. We recall that, for large system sizes, $I_m\sim 1/N$ for delocalized states; any slower scaling is considered a signal of localization. Given the rescaling on the vertical axis of Figure \ref{fig:ipr_increase}, we can conclude that damage induced localization is a robust phenomenon in HFN. The robustness of our results is further confirmed in Figure \ref{fig:ipr_increase} (right), where the above results are averaged over different network realizations, showing little variation with respect to the single realization in Figure \ref{fig:ipr_increase} (left).   

\section{Hausdorff dimension}
\begin{figure}
\begin{center}
\vspace{-0cm}
\includegraphics[width=9cm]{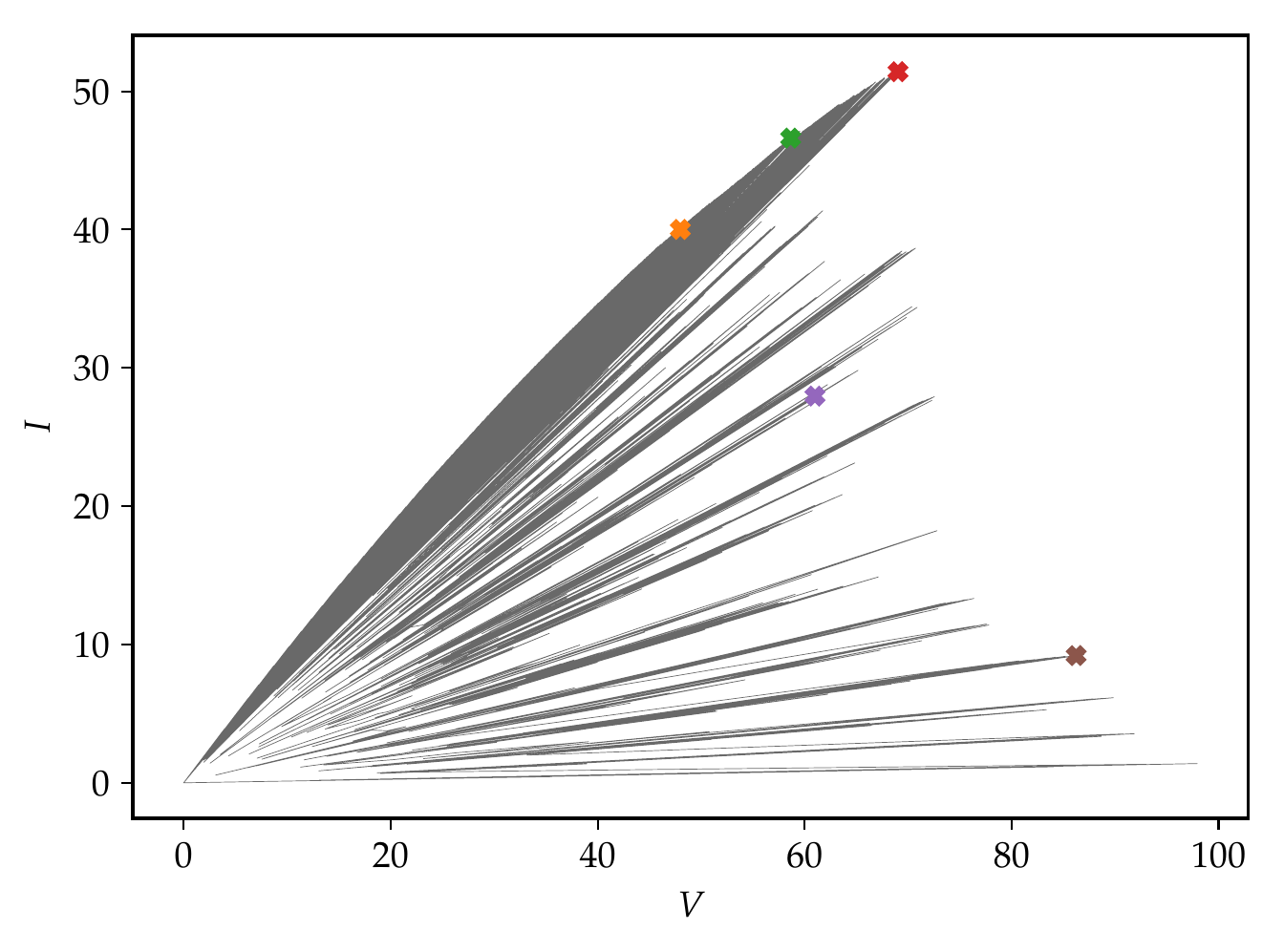}\\
\hspace{0.5cm}\includegraphics[width=10cm]{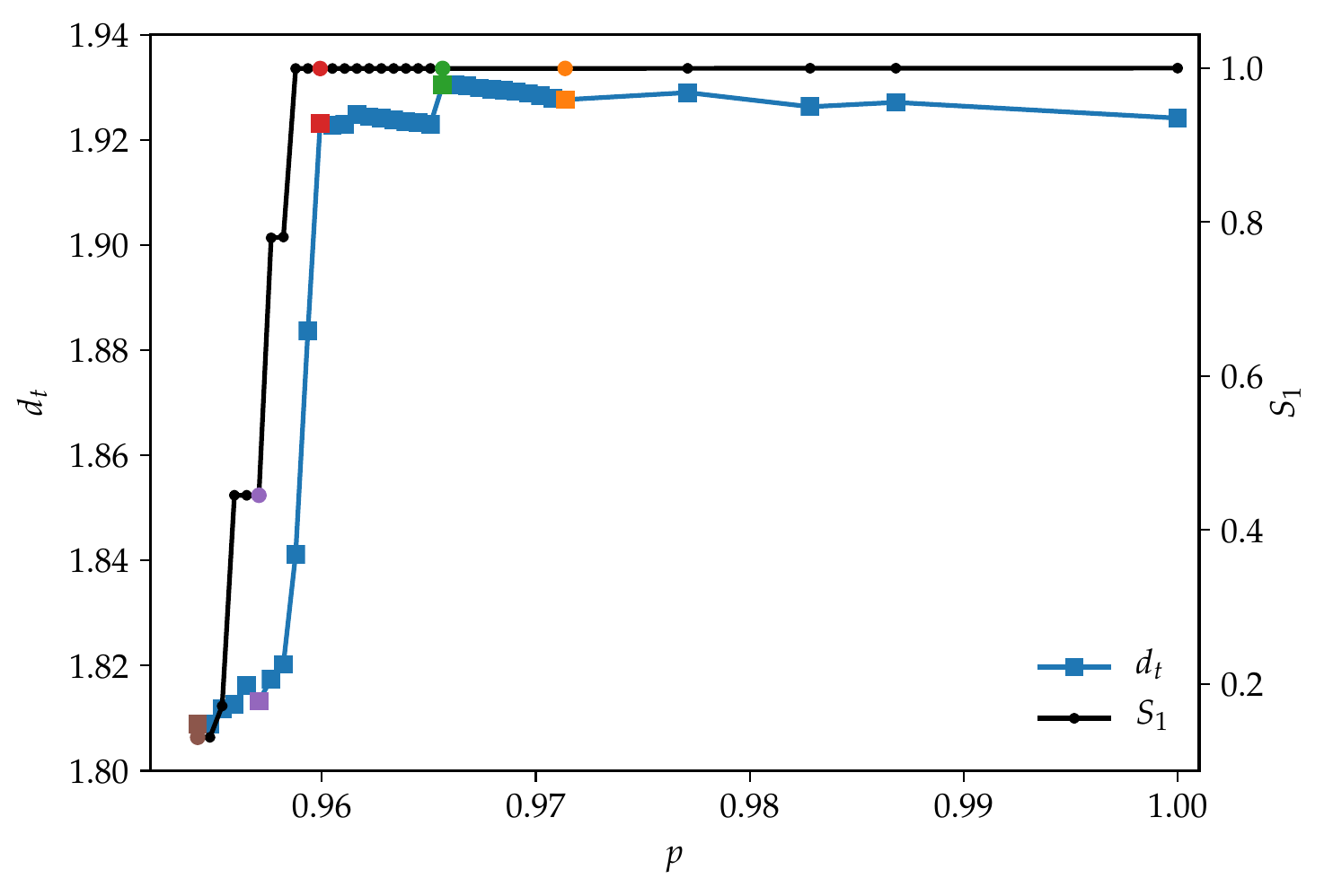}
\caption{Evolution of topological dimension in large-scale HFN under load. Top: reference IV curve, with specific configuration highlighted using different colors. Bottom: topological dimension $d_t$ vs. fraction of burned edges $p$. The size of the largest connected components is also shown for comparison. Different configurations follow the same color scheme as in the top panel, the peak current being marked by a red symbol. Curves should be read from right (undamaged system) to left (cracked system). System size is limited to the largest system available ($L=1024$, $N\approx 10^6$), as dimension measurements are less precise for smaller sizes.}
\label{fig:dimension}
\end{center}
\end{figure}

We have seen above how a sharp increase in eigenvector localization signals failure in HFNs, in a more reliable way than standard methods based on avalanche statistics. More importantly, this type of indicator is a structural one, that is one that can be inferred from the morphology of a sample, rather than from its response to load. The network dimension is in our view another candidate damage indicator, which may exhibit significant changes as failure advances. The dimension of a network is a generalization of the concept of lattice dimension. As various definitions exist, we focus here on the intrinsic Hausdorff dimension $d_t$, or topological dimension, which can be computed as follows: i) for each node $i$ in the network, compute the number $n_{ir}$ of nodes that are $r$ steps away from $i$, for all possible $r$; ii) take the node average $\bar{n}_r$; iii) if $\bar{n}_r\sim r^{d_t}$, $d_t$ is the topological dimension of the network. It is apparent that in regular lattices with nearest neighbor connections $d_t$ equals the lattice dimension, up to finite size effects (the rigorous definition of $d_t$ implies an infinite size limit). It is also clear that any damage process on any network structure can only decrease its value of $d_t$ (up to fluctuations due to the limited accuracy of exponent measurements). Figure \ref{fig:dimension} shows the evolution of $d_t$ for increasing damage in a HFN, throughout the entire loading curve. The peak current is accompanied by a sharp decrease in $d_t$. For comparison, we also show the evolution of $s_1$, the size of the largest connected component, which is normally monitored in percolation theory. As expected from the above considerations, $s_1$ is only mildly related to the softening of the system, and in particular it overestimates the peak current. The topological dimension $d_t$, instead, matches the value of $I_p$ with greater precision.

\section{Discussion and conclusions}
While network-related methods are usually of limited relevance in the study of materials, they become more relevant in the case of materials with complex microstructres, which can be represented by complex contact patterns. This appears to be the norm in biological materials, but also may be relevant in architectured and porous materials, as well as in metamaterials. Such systems can be naturally described as complex network structures. The understanding of the mechanical properties of such materials requires the introduction of quantities whose evolution under damage can be monitored, providing meaningful indicators of failure. In this work we focused on the case of materials with hierarchical microstructures and their modeling as hierarchical fuse networks. Damage in these systems has a peculiarity: it is poorly described by theories that rely on phase transitions and critical points, since scale invariant behavior is observed at any loading stage. Here we focused instead on structural features of the networks at hand, and highlighted their relationships with the damage state of a system under load. We find that the peak current in a HFN is accompanied by a sharp increase in eigenvector localization and a decrease in dimensionality. While both tendencies are easily to understand from an intuitive viewpoint, we find that their onsets mark precisely the peak current values in our simulations. 

\begin{figure}
\begin{center}
\vspace{-0cm}
\includegraphics[width=9cm]{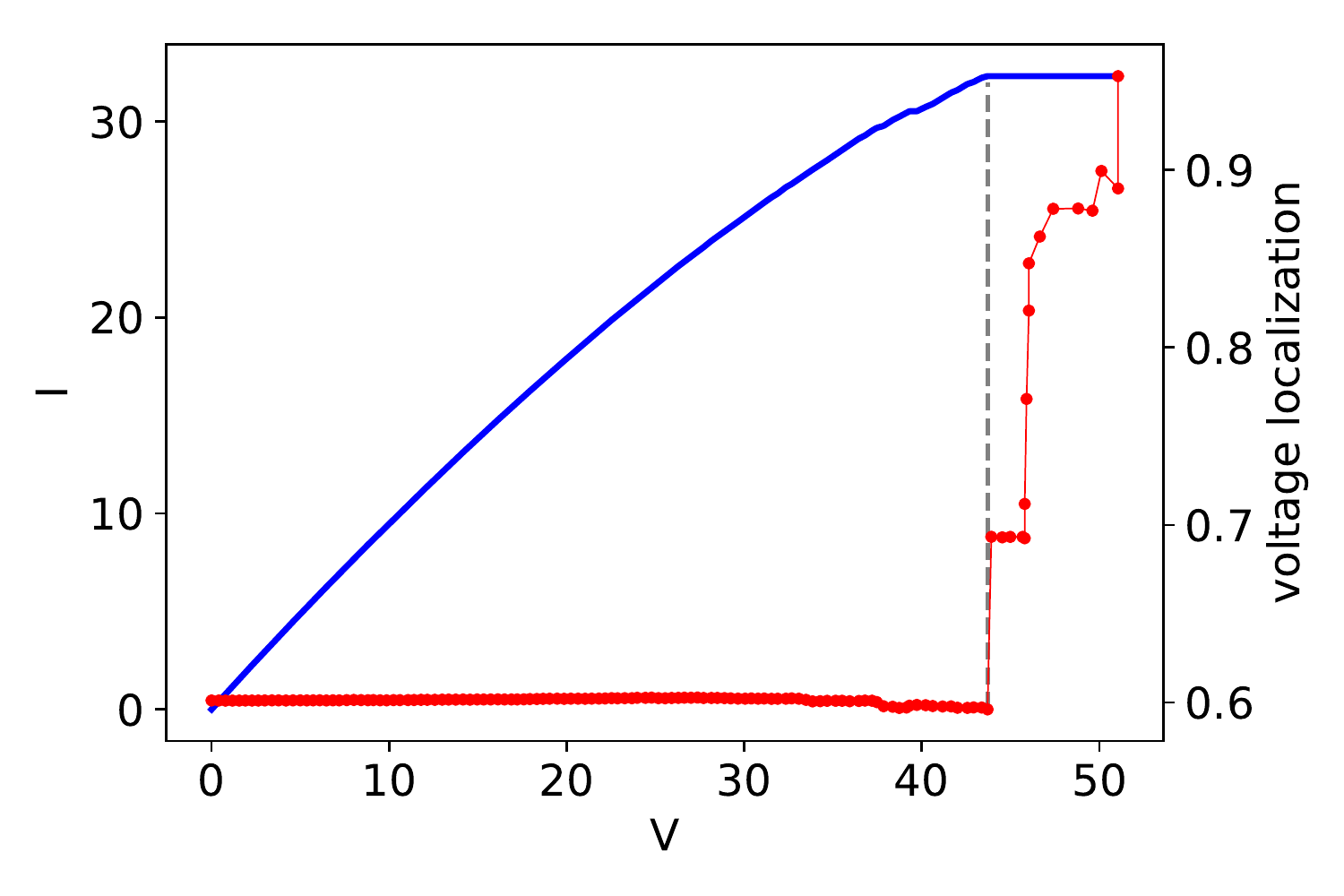}
\caption{Simulation of a current control deformation protocol, for a HFN of size $L=512$. The $I-V$ curve and the inverse participation ratio of the voltage profile are shown. The dashed grey line indicates the peak current state.}
\label{fig:voltage_localization}
\end{center}
\end{figure}

Our results provide general guidelines regarding what spectral and topological properties materials should possess in order to recover the behavior of HFN. A question of more practical relevance would be how these properties can be measured in experiments. The gradual shift of low eigenvalues to lower values for increasing damage is probably the easiest to measure, as each eigenvalue $\lambda_i$ of the lattice Laplacian is proportional to the square of the $i-$th lowest oscillation frequency, $\omega_i^2$. It is also easy to show that in the case of over-damped relaxation, $\lambda_i$ is inversely proportional to the relaxation time of the $i-$th longest relaxation time $\lambda_i\sim 1/\tau_i$. These quantities are amenable to direct measurement e.g. in terms of acoustic spectral properties of a component or structure. As for the possibility to measure eigenvectors, Eq. \ref{eq:solution} states that the voltage field is approximately a linear combination of the eigenvectors in the lower spectral edge: the localization properties of individual eigenvectors should carry over to voltages (in terms of mechanical analogue: local strains) and, by extension, to any property whose spatial structure is governed by the Laplacian operator. To visualize this, Figure \ref{fig:voltage_localization} shows results from a proof-of-concept study, in which we simulate current-controlled load in a HFN and we quantify the localization of the strain profile at each loading stage. Localization is measured as the inverse participation ratio of the vector $\mathbf{v}$. As the system reaches the peak current and softens, strain localization increases substantially, and this localization may be experimentally accessible e.g. from an analysis of digital image correlation data.        

Beyond linear properties, localized features also emerge within inherently non-linear processes associated with patterning and damage in multi-scale systems, with examples ranging from activity patterns in brain networks \cite{Odor2015,Safari2017_NJP} and pattern retrieval in artifical neural networks \cite{Agliari2015_PRL}, to shear band formation in plastically deforming amorphous materials \cite{Sandfeld2014_JSTAT}, or detachment and frictional sliding of bio-mimetic hierarchically patterned materials \cite{Costagliola2016_PRE}. A generic approach to identify damage-induced changes in connectivity patterns associated with some observable $A$ might be to discretize measured spatial patterns of $A$ and compute the associated correlation matrix $C_{ij}$ which may then be appropriately sparsified by applying a physically motivated threshold value. The resulting sparse matrix can be interpreted as the adjacency matrix of a correlation network (a functional network, in the neuroscience jargon), whose spectral properties and topological dimension depend on the state of damage and can be measured as presented in this paper. We hope that the main concepts discussed here can stimulate further interest in the usage of this class of methods to measure damage in broader classes of materials and related to a wider range of structural or functional materials properties.


\section*{Acknowledgements}  
  
The authors acknowledge support from DFG under grants ZA 171/9-1, MO 3049/1-1 and 3049/3-1, as well as by the European commission under H2020-MSCA-RISE project 734485 FRAMED.

\section*{References}
\bibliographystyle{iopart-num}
\providecommand{\newblock}{}

\end{document}